\documentclass[aps,prl,twocolumn,amsfonts,showpacs,superscriptaddress]{revtex4} 
\usepackage{bm}
\usepackage{epsfig,amsopn}
\usepackage{graphicx,color}
\usepackage{amsmath,amssymb, amsfonts, amsthm}
\usepackage{braket}
\newcommand{\sectionprl}[1]{{\par\it #1.---}}
\DeclareMathOperator*{\argmin}{arg\,min}
\DeclareMathOperator*{\argmax}{arg\,max}

\newcommand{\Tr}[0]{ \textrm{Tr}}
\newtheorem{theorem}{Theorem}
\newtheorem{lemma}{Lemma}

\newcommand{\eq}[1]{\begin{equation} #1 \end{equation}}
\newcommand{\eqa}[2]{\begin{equation} #1 \label{#2} \end{equation}}
\newcommand{\balign}[1]{\begin{align} #1 \end{align}}

\newcommand{\todayd}{\the\year/\the\month/\the\day}

\newcommand{\lb}{\label}
\newcommand{\nt}{\notag}

\newcommand{\eref}[1]{Eq.~\eqref{#1}}

\newcommand{\bpf}[1]{\begin{proof} #1 \end{proof}}

\newcommand{\bel}{\begin{easylist}}
\newcommand{\eel}{\end{easylist}}

\def \({\left(}
\def \){\right)}
\def \[{\left[}
\def \]{\right]}

\newcommand{\abs}[1]{\left|#1\right|}

\newcommand{\sumtwo}[2]%
{\mathop{\sum_{#1}}_{#2}}
\newcommand{\sumthree}[3]%
{\mathop{\mathop{\sum_{#1}}_{#2}}_{#3}}
\newcommand{\sumfour}[4]%
{\mathop{\mathop{\mathop{\sum_{#1}}_{#2}}_{#3}}_{#4}} 
\newcommand{\prodtwo}[2]%
{\mathop{\prod_{#1}}_{#2}}
\newcommand{\mintwo}[2]%
{\mathop{\min_{#1}}_{#2}}
\newcommand{\maxtwo}[2]%
{\mathop{\max_{#1}}_{#2}}
\newcommand{\maxthree}[3]%
{\mathop{\mathop{\max_{#1}}_{#2}}_{#3}}
\newcommand{\limtwo}[2]%
{\mathop{\lim_{#1}}_{#2}}
\newcommand{\suptwo}[2]%
{\mathop{\sup_{#1}}_{#2}}
\newcommand{\supthree}[3]%
{\mathop{\mathop{\sup_{#1}}_{#2}}_{#3}}
\newcommand{\supfour}[4]%
{\mathop{\mathop{\mathop{\sup_{#1}}_{#2}}_{#3}}_{#4}} 
\newcommand{\inftwo}[2]%
{\mathop{\inf_{#1}}_{#2}}
\newcommand{\infthree}[3]%
{\mathop{\mathop{\inf_{#1}}_{#2}}_{#3}}
\newcommand{\inffour}[4]%
{\mathop{\mathop{\mathop{\inf_{#1}}_{#2}}_{#3}}_{#4}} 

\newcommand\calE{{\cal E}}

\newcommand\calT{{\cal T}}

\newcommand{\Brakket}[3]{\left\langle #1 \middle|#2 \middle| #3 \right\rangle}

\begin{document}
\title{Uncertainty relations in implementation of unitary operations}
\author{Hiroyasu Tajima}
\affiliation{Department of Communication Engineering and Informatics, University of Electro-Communications, 1-5-1 Chofugaoka, Chofu, Tokyo, 182-8585, Japan}

\author{Naoto Shiraishi}
\affiliation{Department of Physics, Keio University, 3-14-1 Hiyoshi, Yokohama, 223-8522, Japan} 

\author{Keiji Saito}
\affiliation{Department of Physics, Keio University, 3-14-1 Hiyoshi, Yokohama, 223-8522, Japan}

\begin{abstract}
The underlying mechanism in the implementation of unitary operation on a system with an external apparatus is studied. 
We implement the unitary time evolution in the system as a physical phenomenon that results from the interaction between the system and the apparatus. 
We investigate the fundamental limitation of an accurate implementation for  the desired unitary time evolution. 
This limitation is manifested in the form of trade-off relations between the accuracy of the implementation and quantum fluctuation of energy in the external apparatus. Our relations clearly show that an accurate unitary operation requires large energy fluctuation inside the apparatus originated from quantum fluctuation.
\end{abstract}


\maketitle
\sectionprl{Introduction}
Recent technological developments have realized elaborate quantum manipulation on a microscopic level with high accuracy. 
In construction of quantum information devices including quantum computers, experimental techniques for qubit control have been intensively studied, and nontrivial quantum manipulation is realized ~\cite{nakamura1999, barends2013, song2017}. Another important example is quantum heat engines, in which a small quantum system such as a single atom is thermodynamically operated~\cite{cottet2017, lab1, dorner2014, lab2, masuyama2017}. Accurate unitary dynamics in such a deep quantum regime are realized by developing sophisticated experimental apparatus that controls system's parameters.

\begin{figure}
\includegraphics[width=6.5cm]{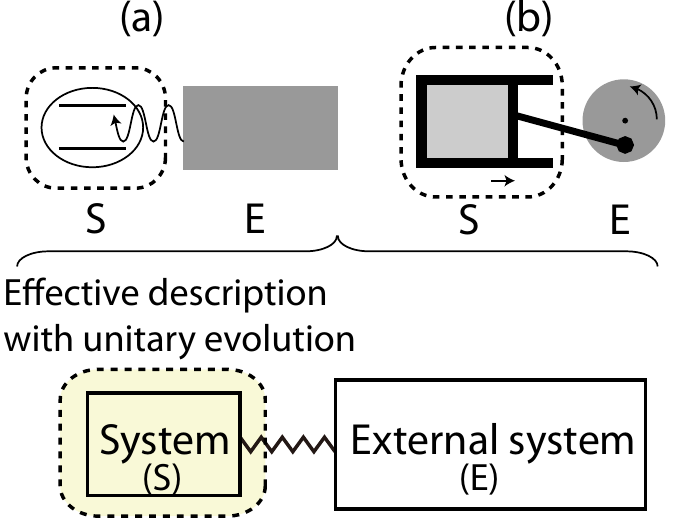}
\caption{Schematic examples of implementation of unitary operation by the experimental apparatus. (a): a qubit system controlled by the electromagnetic field. (b):  a heat engine controlled by a moving piston. The picture shown below is a general composite model (S+E) to study the mechanism of unitary time evolution in the system (S).}
\label{fig1}
\end{figure}

Let us consider the implementation of some unitary transformation on the system.
Let $\rho_S$ and $\tilde{H}_S (t)$ respectively be the initial density matrix of the system and the time-dependent Hamiltonian that leads to the desired unitary operation.
Then, the density matrix at time $\tau$ is given by the unitary transformation $V_S\rho_SV^\dagger_S$ with the unitary operator 
\begin{align}
V_S &:=\calT \exp \left( -i\int_0^\tau dt\, \tilde{H}_S(t) \right) \, ,  \label{us}
\end{align}
where ${\cal T}$ represents the time-ordered product and $\hbar$ is set to unity. 
To implement this unitary transformation as a physical phenomenon, 
we employ an external apparatus and make it physically interact with the system.
Then, the desired unitary transformation $V_S$ is physically realized as a dynamics of a composite system of the system and the external apparatus.
See Fig.\ref{fig1}(a) and (b) for schematic examples.

This setup is generically described by the composite system of system (S) and the external system (E) depicted in the lower figure of Fig.\ref{fig1}.
The simplest example showing such a  realization of a unitary time evolution is the Jayes-Cumming model, which is a model for the cavity QED \cite{percell1946,jaynes1963}. In the cavity QED, a single atom interacts with photons in a cavity. In our setup, the atom and the cavity mode correspond to the system and the external system, respectively. In the classical field limit, the dynamics of the atom is given by the unitary time evolution with the time-dependent Hamiltonian under classical electromagnetic fields. 
Other important examples can be seen in studies on the autonomous heat engines~\cite{catalyst, Brandao2013, Malabarba2015, Woods2016}. Especially, \r{A}berg proposed an idea of autonomous implementation of unitary operation of a system by attaching an external system that has unbounded energy levels with constant energy spacing~\cite{catalyst}. 
However, most previous studies using the setup of composite systems have treated specific models, and thus the general pictures for unitary time evolution independent of the models have remained unclear. 


Motivated by this background, in this letter, we investigate a general picture for the implementation of the unitary time evolution.
We here focus on two quantities: 
The first quantity stands for a distance between the actual system's dynamics and desired unitary time evolution, and the second one is the energy fluctuation of the external system. 
We derive uncertainty type inequalities between these two quantities which capture a fundamental limitation on the implementation of unitary operation.
In particular, these inequalities show that realizing perfectly a desired unitary dynamics and vanishing energy fluctuation in an external system are incompatible. 
In addition, we show that the energy fluctuation must have quantum origin, i.e., as an initial state in the external system, a superposition of many energy eigenstates with a broad energy spectrum is necessary to realize a unitary transformation with high accuracy.

\sectionprl{Setup and first uncertainty relation}
Consider a quantum system $S$ whose Hilbert space and Hamiltonian are ${\cal H}_S$ and $H_{S}$, respectively.
We set the lowest energy of the system to zero and assume that all other eigenergies are finite.
Let us try to implement some unitary transformation $U_S$ on $S$.
To this end, we consider the following steps:

\vspace{5pt}

\textit{Step 1:} We prepare an external quantum system $E$, whose Hilbert space and Hamiltonian are denoted by ${\cal H}_E$ and $H_{E}$, respectively. 
We set the initial state of $E$ as $\sigma_E$.

\textit{Step 2:} We perform an energy preserving CPTP-map (Completely Positive and Trace Preserving map) $\Lambda_{SE}$ on the composite system $SE$.
Then, for the initial state of the system $\rho_S$, the time evolution of the system is written as follows:
\begin{align}
\Lambda_{S}(\rho_S):=\Tr_{E}[\Lambda_{SE}(\rho_S\otimes\sigma_E)].
\end{align}

\vspace{5pt}

For simplicity, as the CPTP map we confine ourselves to consider the unitary transformation described by the following time-independent Hamiltonian \cite{footnoteA}: 
\begin{align}
\Lambda_{SE}(\rho):=e^{-iH\tau}\rho e^{iH\tau},\enskip H &= H_S + H_{SE} + H_E \, , \label{hamil}
\end{align}
where $H_{SE}$ is the interaction Hamiltonian between $S$ and $E$.
We also assume that the energy $H_S+H_E$ is conserved and $[H_S+H_E,e^{-iH\tau}]=0$ is satisfied.
Remark that the latter condition can be loosened, which will be discussed  later.

For given $H_S$ and the initial state of the system $\rho_S$, the actual time evolution of the system, $\Lambda_S(\rho_S)$, is determined by the external system $({\cal H}_E, H_E)$, its initial state $\sigma_E$, the interaction $H_{SE}$ and the time $\tau$.
Hence, the set ${\cal I}:=({\cal H}_E, H_E, \sigma_E, H_{SE}, \tau)$ specifies the implementation of $U_S$~\cite{footnoteB}.
Therefore, we hereafter call the set ${\cal I}$ the ``implementation set'' of $U_S$.
When $\Lambda_S (\rho_S)$ approximates $U_S \rho_S U_S^{\dagger}$ accurately for arbitrary initial density matrix $\rho_S$, we regard that ${\cal I}$ is a good set for the implementing $U_S$.
The aim of this letter is to clarify inevitable limitations on such ``good'' implementation sets ${\cal I}$ for desired $U_S$.

Let us introduce the degree of accuracy of approximation between the actual time evolution and the desired unitary evolution. 
We quantify this by the maximum distance between the final state of the actual time evolution $\Lambda_S(\rho_S)$ and that of the desired unitary time evolution $U_S \rho_S U_S^{\dagger}$:
\begin{align}
\delta_U &:= \left[ \max_{\rho_S} L_B (\Lambda_S(\rho_S) \, , \, U_S \rho_S U_S^{\dagger} \,)\right]^{1/2} \, , \label{deltac}
\end{align}   
where $L_B (\rho_1 , \rho_2)$ is the Bures distance between the states $\rho_1$ and $\rho_2$ defined as \cite{nielsen,hayashi}
\begin{align}
\begin{split}
L_B (\rho_1 , \rho_2) &:= \left[ 1 - F (\rho_1 , \rho_2 )\right]^{1/2}  \, ,  \\
 F (\rho_1 , \rho_2 ) &:= {\rm Tr} \sqrt{\sqrt{\rho_1} \rho_2 \sqrt{\rho_1} } \, .
\end{split}
\end{align}
Here, $F (\rho_1 , \rho_2 )$ is the quantum fidelity for the density matrices $\rho_1$ and $\rho_2$. 
A large $\delta_U$ implies that the description with $\Lambda_S$ fails to approximate the desired unitary $U_S$.
Note that if $\rho_S$ is a pure state while $\Lambda_S(\rho_S)$ has low purity, the quantity $\delta_U$ inevitably becomes large.  
Since our interest is in the good implementation sets, we restrict our attention to the small $\delta_U$ regime.

We also introduce the energy fluctuation of the initial state of the external system defined as
\begin{align}
\delta_E &:= \left[\langle ( \, H_E - \langle H_E \rangle_E \, )^2 \rangle_E \right]^{1/2} \,  . \label{deltae}
\end{align}
Here, $\langle ... \rangle_E$ is the average over the initial density matrix in the external system; $\langle ... \rangle_E:={\rm Tr}_E [... \sigma_E]$. 

We now explain the uncertainty relation in implementation of a unitary time evolution. 
We consider the regime of implementations with high accuracy.
Namely, we consider the implementation sets of small $\delta_U$ satisfying $\delta_U < \|[H_{S},U_{S}]\|_{}/(40 \| H_S \|_{})$, where $\|A\|_{}$ is the spectral norm of an operator $A$. 
In this region, any implementation set ${\cal I}$ satisfies the following trade-off relation between $\delta_E$ and $\delta_U$:
\begin{align}
\delta_E \delta_U \ge \frac{\|[H_S,U_S]\|_{}}{40} \, . \label{tradeoff}
\end{align}
This is our first main result.
The norm of the commutator $\|[H_{S},U_{S}]\|_{}$ is equivalent to the maximum change in energy of the system:
\begin{align}
\|[H_{S},U_{S}]\|_{} &= \max_{\rho_S} \left|\, {\rm Tr} [ H_S (\rho_S-U_S\rho_SU^{\dagger}_S)]   \,\right|\, .  \label{wmax}
\end{align}
We provide the outline of the derivation of (\ref{tradeoff}) later. 
The key observation in the derivation is that to implement the unitary time evolution with high accuracy, the state in the external system must be less affected by the system's energy change in time. 
We will show this key observation in two inequalities \eqref{lemma1} and \eqref{lemma2}.
The relation (\ref{tradeoff}) concludes that a {\it large initial energy fluctuation is necessary to implement the unitary operation when the desired operation changes energy in the system.} 
From the relation (\ref{tradeoff}), in general, the perfect implementation of unitary operation and vanishing energy fluctuation in the external system are incompatible. The only exception is the case involving no energy change in the system, where $\|[H_{S},U_{S}]\|_{}=0$. In this case, we can always give a proper ${\cal I}$ satisfying $\delta_E=0$ and $\delta_U=0$ at the same time. 

\sectionprl{Second uncertainty relation}
Our first inequality \eqref{tradeoff} does not specify the origin of the energy fluctuation in $\sigma_E$, and thus it does not distinguish large energy fluctuation caused by the classical mixture and that by the quantum superposition of many energy eigenstates.
However, many studies on the open quantum systems have shown that the classical mixture in the external system leads to not unitary but dissipative dynamics of the system, even if the energy variance is large~\cite{openquantum}. 
This implies that to implement the unitary time evolution with high accuracy, the origin of the energy fluctuation in the external system should be a quantum superposition, not a classical mixture. To confirm this, we derive the second uncertainty relation, which is related to the quantum superposition in the initial state. To this end, we express the initial state in the following form:
\begin{eqnarray}
\sigma_E = \sum_j p_j |\phi_{E,j} \rangle \langle \phi_{E,j} |
 \, , \label{sigmaec}
\end{eqnarray}
Note that there may be arbitrariness of decompositions $\{p_j, |\phi_{E,j}\rangle \}$ for fixed $\sigma_E$, including the case of a non-orthogonal set of $\{ |\phi_{E,j} \rangle \}$. We define a quantity that measures the energy fluctuation in the form of a quantum superposition
\begin{eqnarray}
\delta_{EQ} := \min_{\{ p_j, |\phi_{E,j}\rangle \} \atop {\rm fixed}\,\sigma_E}\! \left[ \sum_j p_j \langle ( H_E - \langle H_E \rangle_{\phi_{E,j} } )^2 \rangle_{\phi_{E,j} } \right]^{1/2}\!\!\!\!\!, \enskip
\end{eqnarray} 
where $\langle ...\rangle_{\phi_{E,j}}:= \langle \phi_{E,j} | ... | \phi_{E,j} \rangle$ and we take the minimum of all possible decompositions $\{p_j,|\phi_{E,j}\rangle \}$ for a given $\sigma_E$.
If the origin of the fluctuation $\delta_E$ is completely classical, namely if all of $| \phi_{E,j} \rangle $ are energy eigenstates of the Hamiltonian $H_E$, the quantity $\delta_{EQ}$ is exactly zero. 
The finiteness of $\delta_{EQ}$ requires that $| \phi_{E,j} \rangle$ is a superposition of energy eigenstates with different energy.
In particular, if $\sigma_E$ is a pure state, the quantity $\delta_{EQ}$ is equal to $\delta_E$.
Therefore, $\delta_{EQ}$ can be interpreted as a measure of the energy fluctuation with a quantum origin.
Also, it is known that the quantity $\delta_{EQ}$ is equivalent to the quantum Fisher information \cite{Q-Fisher1,Q-Fisher2}.

As the second main result in this letter, we show the following uncertainty relation between $\delta_U$ and $\delta_{EQ}$
\begin{eqnarray}
\delta_{EQ} \delta_U \ge \frac{\|[H_{S},U_{S}]\|_{}}{81} \, , \label{tradeoff2}
\end{eqnarray}
for a regime $\delta_U < \|[H_{S},U_{S}]\|_{}/(64 \| H_S \|_{})$. The derivation of this relation is very similar to the first relation (\ref{tradeoff}), but it is lengthy, and we therefore present it in the supplemental material \cite{supple}. The inequality (\ref{tradeoff2}) concludes that {\it the mixed state composed of energy eigenstates cannot realize unitary time evolution, and a superposition of energy eigenstates with a broad energy spectrum in the external system is necessary to implement the unitary operation with high accuracy.} 
Remarkably, the relations (\ref{tradeoff}) and (\ref{tradeoff2}) are valid for any type of external system, and thus they are applicable to specific models including Jayes-Cummings model and a model in Ref.\cite{catalyst}.

\sectionprl{Toy example with high quantum coherence}
We now consider a toy model to obtain better intuition. We consider the Jaynes-Cummings model, which is a composite system of a single qubit and a free photon. The Hamiltonians are given by 
\begin{align}
H_S &= \epsilon (\sigma_z + {\bm 1}) \, ,  ~
H_{SE} = \lambda(\sigma_+ b + b^{\dagger} \sigma_- )\, , ~
H_E = 2\epsilon b^{\dagger} b \, ,
\end{align}
where $\lambda$ is the amplitude of the interaction. The operator $\sigma_{z}$ is the $z$-component of the Pauli matrix, and $\sigma_+$ ($\sigma_-$) flips the spin from down(up) to up(down). The operator $b$ and $b^\dagger$ are the annihilation and creation operators of the boson, respectively. We set the coherent state to the initial state of the external system:
\begin{align}
\sigma_E &= |\alpha \rangle \langle \alpha | \, , ~~
|\alpha \rangle = e^{\alpha (b^\dagger - b) } | 0 \rangle \, , 
\end{align}
where $|0 \rangle$ is the vacuum state, and the parameter $\alpha$ is a real number. If we impose the condition $\lambda\to +0$ with $\lambda \alpha$ set to a constant, the dynamics of the reduced density matrix of the system is exactly described by the unitary time-evolution \cite{supple}, i.e., $\rho_S' = U_S \rho_S U_S^{\dagger}$ where
\begin{align}
\begin{split}
U_S &= {\cal T}e^{ -i \int_0^{\tau} \!\! dt  \, \tilde{H}_{S} (t)}= e^{-i \tau \epsilon \sigma_z } e^{-i\tau \alpha \lambda \sigma_x}  \, , ~\\
\tilde{H}_{S} (t)  &= \epsilon \sigma_z + \lambda \alpha (\sigma_+ e^{-i2\epsilon t} + \sigma_- e^{i 2\epsilon t}) \, .~
\end{split}
\end{align}
The initial energy fluctuation is exactly given by $\delta_E = 2\alpha \epsilon$. For a very large $\alpha$, the photon state is almost a classical state that is driven solely by the Hamiltonian $H_E$. Hence, in the large $\alpha$ limit, the time evolution of the system is described by the effective time-dependent Hamiltonian $\tilde{H}_S (t)$. For a finite $\alpha$, we expect that the description with $\tilde{H}_S (t)$ is imperfect, but the relation (\ref{tradeoff}) is satisfied. 

\begin{figure}
\includegraphics[width=7.5cm]{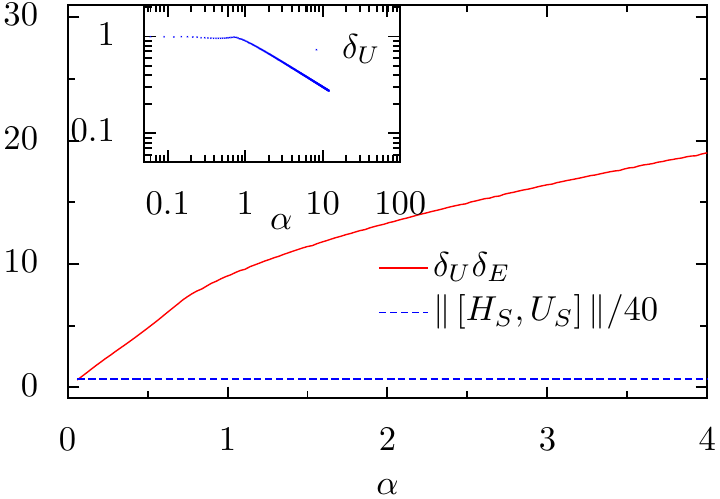}
\caption{
Demonstration of the first uncertainty relation in the Jaynes-Cummings model. The inset shows $\delta_U$ as a function of the parameter $\alpha$. Parameters: $\epsilon=10, \alpha\lambda\tau=\pi/2$ from which one gets $\|[H_{S},U_{S}]\|_{}=2 \epsilon$. The relation \eqref{tradeoff} is justified for $\alpha > \alpha_c$, where we estimate $\alpha_c \sim 500$ from the inset. We showed the data for a numerically computable regime of $\alpha$, which is much smaller than $\alpha_c$. Nevertheless, the relation (\ref{tradeoff}) is satisfied.}
\label{fig3}
\end{figure}  

In Fig.\ref{fig3}, we numerically demonstrate that the uncertainty relation is satisfied in this model. We first generated more than $10^4$ density matrices for $\rho_S$ randomly, and we computed $\delta_U$ within this sampling. The inset of Fig.\ref{fig3} shows a plot of $\delta_U$ as a function of $\alpha$, which clearly shows that the unitary time evolution gives a better description as $\alpha$ increases. Because it is difficult to compute the cases of large $\alpha$, we present data for a numerically available regime. In our proof for the present parameter set, the uncertainty relation is justified only for the regime of $\alpha > \alpha_c$, where $\alpha_c \sim 500$. Nevertheless, Fig.\ref{fig3} clearly shows that the first uncertainty relation is satisfied, even in the regime of small $\alpha$. Thus, in this example, the condition $\delta_U\le\|[H_S,U_S]\|_{}/(40\|H_S\|_{})$ is much too strong, and our inequality is satisfied beyond the regime.

\sectionprl{Outline of derivation of \eqref{tradeoff}}
Here, we show the outline of the derivation of (\ref{tradeoff}). To prove this, we employ two useful inequalities. The first inequality relates the Bures distance and the variance of any Hermitian operator $A$. We denote the standard deviations of $A$ in a state $\sigma$ by $\delta_{A}(\sigma):=\sqrt{\Tr[A^2\sigma]-\Tr[A\sigma]^2}$. Then, the difference between the expectation value of $A$ for two states $\sigma_1$ and $\sigma_2$ denoted by $\Delta :=|\Tr[A(\sigma_1-\sigma_2)]|$ satisfies~\cite{supple}
\begin{align}
\Delta\le \sqrt{2}L_{B}(\sigma_1,\sigma_2)(\delta_{A}(\sigma_1)+\delta_{A}(\sigma_2)+\Delta). \label{lemma1}
\end{align}
This inequality suggests that if two states $\sigma_1$ and $\sigma_2$ are similar (i.e. small $L_{B}(\sigma_1,\sigma_2)$) but $\Delta$ is large, then at least one of the standard deviations of $A$ in these two states is large. 
The second inequality relates the quantity $\delta_U$ to the final state of the external system. We denote the reduced density matrix of the external system at time $\tau$ with the initial state of the system $\rho_{S, \nu}$ by $\sigma_{E,\nu}' := {\rm Tr}_S \left[ \Lambda_{SE} (\rho_{S,\nu} \otimes \sigma_E )  \right]$, where ${\rm Tr}_S$ is the partial trace with respect to the system. Consider two pure initial states of the system $\rho_{S,\nu_1}$ and $\rho_{S,\nu_2}$, which are orthogonal to each other. Then, for $\delta_U\le1/4$, we have the following inequality~\cite{supple}:
\begin{align}
L_B (\sigma_{E,\nu_1}' ,\sigma_{E,\nu_2}' ) \le  4 \delta_U \, .  \label{lemma2}
\end{align}
This inequality means that if the actual time evolution is close to the unitary time evolution, the final reduced density matrices of the external system starting from different initial states of the system are similar to each other. 
This implies that the states in the external system is robust against the change of initial states of the system.

We consider the two initial density matrices for the system labeled as $\rho_{S,{\rm max}}$ and $\rho_{S,{\rm min}}$, which respectively maximizes and minimizes the energy loss in the system, i.e., 
\begin{equation}
\begin{split}
\rho_{S,{\rm max}}
 &:= \argmax_{\rho_S} {\rm Tr}\left[ ( \rho_S - U_S \rho_S U_S^{\dagger} ) H_S \right], \\
\rho_{S,{\rm min}} &:= \argmin_{\rho_S }{\rm Tr}\left[ ( \rho_S - U_S \rho_S U_S^{\dagger} ) H_S \right] \, .
\end{split}
\label{maxmin}
\end{equation}
Because the matrix $H_S - U_S^{\dagger} H_S U_S$ is a Hermitian matrix, there exist two eigenstates of this Hermitian matrix $\ket{\psi_{\rm max}}$ and $\ket{\psi_{\rm min}}$ ($\braket{\psi_{\rm max}|\psi_{\rm min}}=0$), with which the above two density matrices are expressed as $\rho_{S,{\rm max}}=\ket{\psi_{\rm max}}\bra{\psi_{\rm max}}$ and  $\rho_{S,{\rm min}}=\ket{\psi_{\rm min}}\bra{\psi_{\rm min}}$. In other words, $\rho_{S,{\rm max}}$ and $\rho_{S,{\rm min}}$ are pure and orthogonal to each other. Then, by setting $A=H_E$, $\sigma_1=\sigma_{E,\nu_1}'$ and $\sigma_2=\sigma_{E,\nu_2}'$ in (\ref{lemma1}), and $\rho_{S,\nu_1}=\rho_{S,{\rm max}}$ and $\rho_{S,\nu_2}=\rho_{S,{\rm min}}$ in (\ref{lemma2}), we obtain the inequality 
\eqa{
\Delta \le 4\sqrt{2} \delta_U (2 \tilde{\delta}_E + \Delta ),
}{middle}
where $\tilde{\delta}_E:=\max ( \delta_E (\sigma_{E,{\rm max}}'), \delta_E (\sigma_{E,{\rm min}}' )) $. 

We now consider the high-accuracy regime: $\delta_U \le \|[H_{S},U_{S}]\|_{}/(40 \| H_S \|_{}) <1/40$. 
Using $\delta_U <1/40$ and the inequality \eqref{middle}, we obtain 
\begin{align}
\delta_U \tilde{\delta}_E & \geq \frac{1}{2}\( \frac{1}{4\sqrt{2}}-\frac{1}{40}\) \Delta= \frac{10 - \sqrt{2}}{80\sqrt{2}} \Delta \, . \label{in1}
\end{align} 
Roughly speaking, the quantity $\Delta$ is close to $\|[H_{S},U_{S}]\|_{}$, and the quantity $\delta_E$ is close to $\tilde{\delta}_E$, although there are slight deviations. Rigorous relations for these variables read~\cite{supple}
\begin{align}
\|[H_{S},U_{S}]\|_{} & \le \Delta + 4\sqrt{2}\delta^{2}_U \| H_S \|_{} \, \label{s1}, \\
\tilde{\delta}_E & \le \delta_E + \| H_S \|_{} \, .  \label{s2}
\end{align} 
The combination of Eq.(\ref{s1}) and the condition $\delta_U \le \|[H_{S},U_{S}]\|_{}/(40 \| H_S \|_{})<1/40$ yields $\Delta >(1-\sqrt{2}/400)\|[H_S,U_S]\|_{}$.
By applying the above inequality to the inequality (\ref{in1}), we get
\begin{align}
\delta_U \tilde{\delta}_E  \ge \frac{\|[H_{S},U_{S}]\|_{}}{20} \, . \lb{ul1-last}
\end{align}
Finally, the combination of Eq.(\ref{s2}) and the condition $\delta_U \le \|[H_{S},U_{S}]\|_{}/(40 \| H_S \|_{})$ yields $\delta_U \tilde{\delta}_E \leq \delta_U \delta_E +\|[H_S,U_S]\|_{}/40$. This inequality with \eref{ul1-last} directly implies the uncertainty relation (\ref{tradeoff}).

\sectionprl{Discussion}
In this letter, we considered the underlying mechanism in the implementation of the unitary operation. 
By considering a model of a composite system (Fig.\ref{fig1}), we derived two types of fundamental trade-off relations, i.e., (\ref{tradeoff}) and (\ref{tradeoff2}). 
These relations quantitatively clarified the crucial roles of quantum superposition and a broad energy spectrum in the external system.

Although it is difficult to achieve the equalities in the inequalities (\ref{tradeoff}) and (\ref{tradeoff2}) except for the trivial case $\left[ H_S , U_S\right] =0$, the inequalities explicitly show that the fundamental limitation in the form of the uncertainty type relations actually exist in implementation of unitary operations.
The aim of this letter is to show the existence of a novel type of fundamental limitation.
There is much room to improve the tightness of our inequalities.
In fact, tighter bounds can be derived by using alternatively a more sophisticated but less standard quantifier with the entanglement fidelity \cite{hayashi}.
We explain the results in the supplementary material \cite{supple}.

In our setup, we assumed $[H_S+H_E,e^{-iH\tau}]=0$ in the Hamiltonian for simplicity. However, we can consider wider classes of Hamiltonian by introducing the deviation $\chi:=\|[H_S+H_E,e^{-iH\tau}]\|$\cite{footnoteC}. Then, one can derive the following inequalities \cite{supple}: 
\begin{align}
\delta_{E}\delta_{U}&\ge\frac{\|[U_S,H_S]\|-\chi}{40},  \label{tradeoffe} \\
\delta_{EQ}\delta_{U}&\ge\frac{\|[U_S,H_S]\|-\chi}{81}  \label{tradeoff2e}. 
\end{align}
These inequalities imply that the relations (\ref{tradeoff}) and (\ref{tradeoff2}) are continuously connected to the results for the general coupling form.

Our setup is relevant to quantum heat engines, particularly when one considers a work storage, which is a physical object for storing work~\cite{Popescu2014,Popescu2015,Malabarba,Woods,HO,brandao2015,SSP,TWO,oneshot3,gourreview,T-W,MTH,Tasaki2015}. Applying our theory to problems on quantum coherence and the measurement procedure of quantum work \cite{m-based,Marti} will be an intriguing future research subject. 
We must consider time-evolution in the present argument, and hence, it will be intriguing to consider the relationships between our argument and the other type of trade-off relations in the time domain \cite{Uhlmann1992,Mandelstam1945,Anandan1990,Braunstein1994,Erker2017,Funo2017}.

\acknowledgments
We thank Yasunobu Nakamura, Ken Funo, Shingo Kono, and Yusuke Kinoshita for the fruitful discussion and helpful comments.
The present work was supported by JSPS Grants-in-Aid for Scientific Research No. JP14J07602 (NS), No. JP25103003 (KS), and No. JP16H02211 (KS).

\clearpage

\begin{widetext}
\begin{center}
{\large \bf Supplemental Material for \protect \\ 
``Uncertainty relations in implementation of unitary operation''}\\
\vspace*{0.3cm}
Hiroyasu Tajima$^{1}$, Naoto Shiraishi$^{2}$ and Keiji Saito$^{2}$ \\
\vspace*{0.1cm}
$^{1}${\small \em Center for Emergent Matter Science (CEMS), RIKEN, Wako, Saitama 351-0198 Japan}
 \\
$^{2}${\small \em Department of Physics, Keio University, Yokohama 223-8522, Japan}
\end{center}

\setcounter{equation}{0}
\setcounter{page}{1}
\renewcommand{\theequation}{S.\arabic{equation}}

In this supplementary, we refer to the vector representation of a pure state $\rho$ as $\ket{\rho}$.
We write a product state of two systems $\ket{\phi}\otimes\ket{\psi}$ as $\ket{\phi\otimes\psi}$.

\section{Proof of Eq.(\ref{lemma1})}

In this section, we derive \eqref{lemma1} in the main text, which reappears below:
\begin{align}
\Delta\le \sqrt{2}L_{B}(\sigma_1,\sigma_2)(\delta_{A}(\sigma_1)+\delta_{A}(\sigma_2)+\Delta).
\end{align}

\bpf{
Using eigenstates of $A$ ($A=\sum_{i}a_{i}\ket{i}\bra{i}$), we define the probability distribution $P$ and $Q$ as
\begin{align}
p_{i}&:=\mathrm{Tr}[\ket{i}\bra{i}\sigma_1],\\
q_{i}&:=\mathrm{Tr}[\ket{i}\bra{i}\sigma_2].
\end{align}
The variances of $A$ in $P$ ($Q$) are equal to that in $\sigma_1$ ($\sigma_2$).
It is known that the quantum Bures distance for two quantum states is always larger than the classical Bures distance (the Hellinger distance) for the distributions of the diagonal elements of the two states with any fixed basis \cite{nielsen}, which reads
\begin{align}
L_{B}(P,Q):=\sqrt{1-\sum_{i}\sqrt{p_{i}q_{i}}}\le L_{B}(\sigma_1,\sigma_2).
\end{align}
Noting this relation, we see that the desired relation \eqref{lemma1} follows from the following inequality:
\begin{align}
\Delta\le \sqrt{2}L_{B}(P,Q)(\delta_{A}(P)+\delta_{A}(Q)+\Delta). \lb{mid-lemma1}
\end{align}
This inequality is obtained as a special case of the following inequality for an arbitrary real number $X$:
\begin{align}
\Delta&=|\sum_{i}(p_{i}-q_{i})(a_{i}-X)|\nonumber\\
&=|\sum_{i}(\sqrt{p_{i}}-\sqrt{q_{i}})(\sqrt{p_{i}}+\sqrt{q_{i}})(a_{i}-X)|\nonumber\\
&\le\sqrt{\sum_{i}(\sqrt{p_{i}}-\sqrt{q_{i}})^2}\sqrt{\sum_{i}(\sqrt{p_{i}}+\sqrt{q_{i}})^2(a_{i}-X)^2}\nonumber\\
&=\sqrt{2}L_{B}(P,Q)\sqrt{\sum_{i}(p_{i}+q_{i}+2\sqrt{p_{i}q_{i}})(a_{i}-X)^2}\nonumber\\
&\le \sqrt{2}L_{B}(P,Q)\sqrt{\sum_{i}p_{i}(a_{i}-X)^2+\sum_{i}q_{i}(a_{i}-X)^2+2\sqrt{\sum_{i}p_{i}(a_{i}-X)^2}\sqrt{\sum_{i}q_{i}(a_{i}-X)^2}}\nonumber\\
&= \sqrt{2}L_{B}(P,Q)(\sqrt{\sum_{i}p_{i}(a_{i}-X)^2}+\sqrt{\sum_{i}q_{i}(a_{i}-X)^2}).
\end{align}
In the third and fifth lines, we used the Schwarz inequality.
By substituting $\Tr[\sigma_1 A]$ into $X$, we obtain \eref{mid-lemma1}.
}

\section{Proof of Eq.(\ref{lemma2}) }
The inequality \eqref{lemma2} is derived as a corollary of the following theorem:
\begin{theorem}\label{unify}
Consider a composite system of two quantum systems $A$ and $B$ (see Fig.\ref{kaisetu}).
Let $\rho_{A,\nu_1}$ and $\rho_{A,\nu_2}$ be two arbitrary pure states on $A$, which are orthogonal to each other.
We introduce their superposition denoted by $\rho_{A,\nu_{1+2}}:=(\ket{\rho_{A,\nu_1}}+\ket{\rho_{A,\nu_2}})/\sqrt{2}$.
The initial state of $B$ denoted by $\sigma_{B}$ is supposed to be a pure state.
We consider the unitary time evolution $V_{AB}$ of the composite system from the initial state $\rho_{A,\nu}\otimes\sigma_{B}$ ($\nu=\nu_1,\nu_2,\nu_{1+2}$).
The final states of $A$ and $B$ are written as $\rho'_{A,\nu}:={\rm Tr}_{B}[V_{AB}(\rho_{A,\nu}\otimes\sigma_{B})V^{\dagger}_{AB}]$ and $\sigma'_{B,\nu}:={\rm Tr}_{B}[V_{AB}(\rho_{A,\nu}\otimes\sigma_{B})V^{\dagger}_{AB}]$, respectively.
We fix a unitary operator $U_{A}$ on $A$, and define $\delta_{U,\nu}:=[L_{B}(\rho'_{A,\nu},U_{A}\rho_{A,\nu}U^{\dagger}_{A})]^{1/2}$.
Then, for any $\rho_{A,\nu_1}$, $\rho_{A,\nu_2}$, $V_{AB}$, and $U_A$, the following inequality holds:
\begin{align}
L_{B}(\sigma'_{B,\nu_{1}},\sigma'_{B,\nu_{2}})\le2\sqrt{\delta^{4}_{U,\nu_{1}}+\delta^{4}_{U,\nu_{2}}+\sqrt{2}(\delta^{2}_{U,\nu_{1}}+\delta^{2}_{U,\nu_{2}})}+2\sqrt{2}\delta^{2}_{U,\nu_{1+2}}+\delta^{2}_{U,\nu_{1}}+\delta^{2}_{U,\nu_{2}}\label{u-ineq}.
\end{align}
\end{theorem}

\begin{figure}[b]
\begin{center}
\includegraphics[scale=0.28]{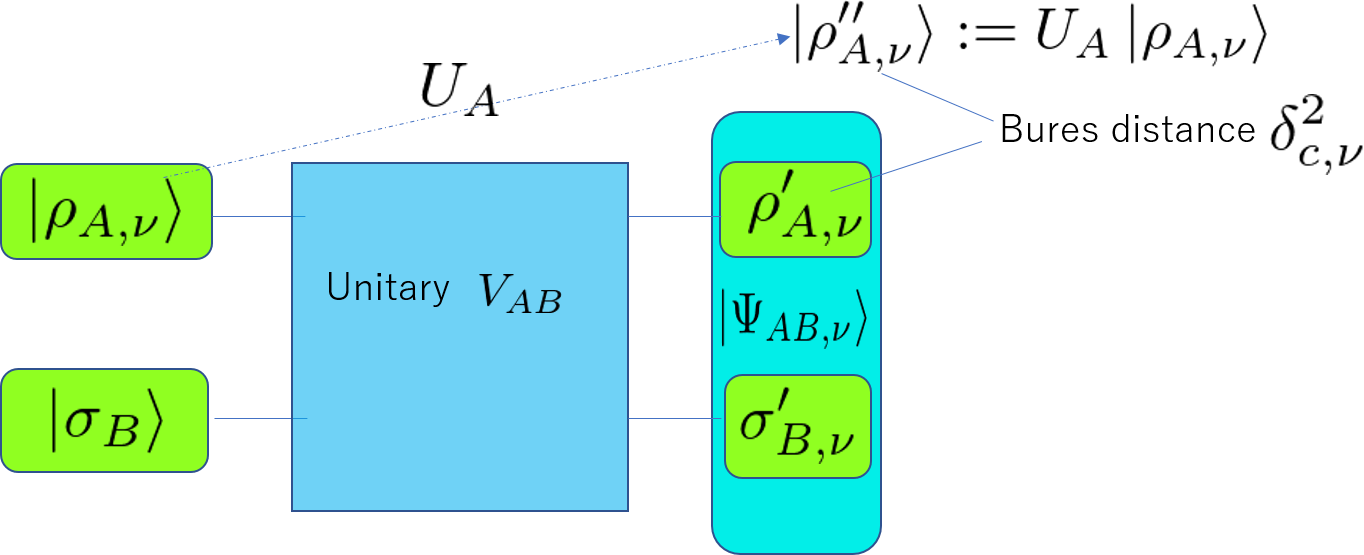}
\end{center}
\caption{A schematic diagram of the setup of Theorem \ref{unify}.
In this supplementary, we refer to the vector representation of a pure state $\rho$ as $\ket{\rho}$.}
\label{kaisetu}
\end{figure}

Before proving this theorem, we first show how \eqref{lemma2} is derived from Theorem \ref{unify}.
As a simple case, we first derive \eqref{lemma2} for the case where $\Lambda_{SE}$ is a unitary dynamics $\Lambda_{SE}$ and $\sigma_{E}$ is a pure state.
In this case, by setting $S$ and $E$ to $A$ and $B$ in Theorem \ref{unify}, the condition $\delta_{U,\nu}\le\delta_{U}\le1/8$ $(\nu=\nu_1,\nu_2,\nu_{1+2})$ suggests \eref{lemma2}
\begin{align}
L_{B}(\sigma'_{E,\nu_{1}},\sigma'_{E,\nu_{2}})\le\left(2\sqrt{\frac{1}{32}+2\sqrt{2}}+\frac{2\sqrt{2}+2}{8}\right)\delta_{U}\le4\delta_{U}.
\end{align}

Next, we consider the general case where $\Lambda_{SE}$ is a general CPTP map, and $\sigma_{E}$ is a mixed state.
Note that an arbitrary CPTP map $\Lambda_{SE}$ can be written as unitary dynamics of the focusing system $SE$ and an extra system $E'$.
In other words, there exists a proper extra system $E'$, an initial state $\rho_{E'}$ of $E'$, and a unitary $V_{SEE'}$ on $SEE'$ such that the CPTP map $\Lambda_{SE}$ is reproduced as
\begin{align}
\Lambda_{SE}(\rho_{S,\nu}\otimes\sigma_{E}):=\Tr_{E'}[V_{SEE'}(\rho_{S,\nu}\otimes\sigma_{E}\otimes\rho_{E'})V^{\dagger}_{SEE'}].
\end{align}
By performing the purification of $\sigma_{E}\otimes\rho_{E'}$, we have a reference system $R$ and a pure state $\sigma_{EE'R}$ such that
\begin{align}
\Tr_{R}[\sigma_{EE'R}]=\sigma_{E}\otimes\rho_{E'}.
\end{align}
Setting $S$ and $EE'R$ to $A$ and $B$ in Theorem \ref{unify}, we obtain
\begin{align}
L_{B}(\sigma'_{EE'R,\nu_{1}},\sigma'_{EE'R,\nu_{2}})\le4\delta_{U}.
\end{align}
Because the Bures distance does not increase using a CPTP map and the partial trace is a CPTP map, the above inequality reduces to \eqref{lemma2}.

\bpf{[Proof of Theorem \ref{unify}]
We refer to the final state of $AB$ as $\ket{\Psi_{AB,\nu}}:=V_{AB}\ket{\rho_{A,\nu}\otimes\sigma_{B}}$.
Because $\rho''_{A,\nu}:=U_{A}\rho_{A,\nu}U^{\dagger}_{A}$ is a pure state, we can define the pure state of $B$ as 
\eq{
\ket{\sigma''_{B,\nu}}:=\frac{\sum_i \ket{\psi_i}\braket{\rho''_{A,\nu}\otimes \psi_i|\Psi_{AB,\nu}}}{ \sum_i \abs{\braket{\rho''_{A,\nu}\otimes \psi_i|\Psi_{AB,\nu}}}^2},
}
where $\{ \psi_i\}$ is a basis of $B$.
Note that the above definition of $\ket{\sigma''_{B,\nu}}$ is well-defined and independent of the choice of the basis $\{ \psi_i\}$.

First, we calculate the Bures distance between $\sigma'_{B,\nu}$ and $\sigma''_{B,\nu}$.
The monotonicity of the Bures distance under a partial trace leads to
\begin{align}
L_{B}(\sigma'_{B,\nu},\sigma''_{B,\nu})\le L_{B}(\Psi_{AB,\nu},\rho''_{A,\nu}\otimes\sigma''_{B,\nu})=L_{B}(\rho'_{A,\nu},\rho''_{A,\nu})=\delta^{2}_{U,\nu}.\label{8.15.4}
\end{align}

Next, we calculate the Bures distance between $\sigma''_{B,\nu_{1}}$ and $\sigma''_{B,\nu_{2}}$.
To do this, we focus on the following quantity
\eq{
\sqrt{1-\Brakket{\sigma''_{B,\nu_{1+2}}}{\frac{\sigma''_{B,\nu_{1}}+\sigma''_{B,\nu_{2}}}{2}}{\sigma''_{B,\nu_{1+2}}}^{1/2}}=L_B\( \sigma''_{B,\nu_{1+2}}, \frac{\sigma''_{B,\nu_{1}}+\sigma''_{B,\nu_{2}}}{2} \) ,
}
which is evaluated as
\balign{
L_B\( \sigma''_{B,\nu_{1+2}}, \frac{\sigma''_{B,\nu_{1}}+\sigma''_{B,\nu_{2}}}{2} \)
&\leq L_{B}\( \sigma'_{B,\nu_{1+2}},\frac{\sigma''_{B,\nu_{1}}+\sigma''_{B,\nu_{2}}}{2}\) +L_B ( \sigma'_{B,\nu_{1+2}},  \sigma''_{B,\nu_{1+2}}) \nt \\
&\leq L_{B}\( \sigma'_{B,\nu_{1+2}},\frac{\sigma''_{B,\nu_{1}}+\sigma''_{B,\nu_{2}}}{2}\) +\delta^{2}_{U,\nu_{1+2}} \nt \\
&\leq L_{B}\left(\ket{\Psi_{AB,\nu_{1+2}}},\frac{\ket{\rho''_{A,\nu_1}\otimes\sigma''_{B,\nu_1}}+\ket{\rho''_{A,\nu_2}\otimes\sigma''_{B,\nu_2}}}{\sqrt{2}}\right) +\delta^{2}_{U,\nu_{1+2}}. 
}
In the first line, we used the triangle inequality.
In the second line, we used \eqref{8.15.4}.
In the third line, we used the monotonicity of the Bures distance through the partial trace.
Owing to the relation $F(\rho,\sigma)=\abs{\braket{\rho|\sigma}}$ for pure states $\rho$ and $\sigma$, the first term of the right-hand side is evaluated as
\balign{
&L_{B}\left(\ket{\Psi_{AB,\nu_{1+2}}},\frac{\ket{\rho''_{A,\nu_1}\otimes\sigma''_{B,\nu_1}}+\ket{\rho''_{A,\nu_2}\otimes\sigma''_{B,\nu_2}}}{\sqrt{2}}\right) \nt \\
=&\sqrt{1-F\left(\frac{\ket{\Psi_{AB,\nu_{1}}}+\ket{\Psi_{AB,\nu_{2}}}}{\sqrt{2}},\frac{\ket{\rho''_{A,\nu_1}\otimes\sigma''_{B,\nu_1}}+\ket{\rho''_{A,\nu_2}\otimes\sigma''_{B,\nu_2}}}{\sqrt{2}}\right)} \nt \\
\leq &\sqrt{1-\( 1-\frac{\delta^{4}_{U,\nu_{1}}+\delta^{4}_{U,\nu_{2}}}{2}-\frac{1}{2}(|\braket{\rho''_{A,\nu_1}\otimes \sigma''_{B,\nu_1}|\Psi_{AB,\nu_2}}| +|\braket{\rho''_{A,\nu_2} \otimes \sigma''_{B,\nu_2}| \Psi_{AB,\nu_1}}| )\) } \nt \\
\leq &\sqrt{\frac{\delta^{4}_{U,\nu_{1}}+\delta^{4}_{U,\nu_{2}}}{2}+\frac{\delta^{2}_{U,\nu_{1}}+\delta^{2}_{U,\nu_{2}}}{\sqrt{2}}} .
}
The transformations in the third and fourth lines are confirmed by applying $|\braket{\rho''_{A,\nu} \otimes \sigma''_{B,\nu}| \Psi_{AB,\nu}}|=1-\delta^4_{U,\nu}$ and $\braket{\rho''_{A,\nu_1}\otimes \sigma''_{B,\nu_1}|\rho''_{A,\nu_2}\otimes \sigma''_{B,\nu_2}}=0$ into the relation that if $\braket{a|b}=0$ and $\braket{a|\tilde{a}}\geq 1-\delta$, then $|\braket{\tilde{a}|b}|\le\sqrt{2\delta-\delta^2}\le\sqrt{2\delta}$.
Combining the above result and the following relation, which comes from $L_B^2\leq 1$,
\begin{align}
\Brakket{\sigma''_{B,\nu_{1+2}}}{\frac{\sigma''_{B,\nu_{1}}+\sigma''_{B,\nu_{2}}}{2}}{\sigma''_{B,\nu_{1+2}}}&\le\frac{(1-L^{2}_{B}(\sigma''_{B,\nu_{1+2}},\sigma''_{B,\nu_{1}}))^2+(1-L^{2}_{B}(\sigma''_{B,\nu_{1+2}},\sigma''_{B,\nu_{2}}))^2}{2}
\nonumber\\
&=1-\frac{L^{2}_{B}(\sigma''_{B,\nu_{1+2}},\sigma''_{B,\nu_{1}})+L^{2}_{B}(\sigma''_{B,\nu_{1+2}},\sigma''_{B,\nu_{2}})}{2}\nonumber\\
&\le1-\frac{(L_{B}(\sigma''_{B,\nu_{1+2}},\sigma''_{B,\nu_{1}})+L_{B}(\sigma''_{B,\nu_{1+2}},\sigma''_{B,\nu_{2}}))^2}{4}\nonumber\\
&\le1-\frac{L^{2}_{B}(\sigma''_{B,\nu_{1}},\sigma''_{B,\nu_{2}})}{4},
\end{align}
we arrive at the desired inequality on $L^{2}_{B}(\sigma''_{B,\nu_{1}},\sigma''_{B,\nu_{2}})$;
\begin{align}
L_{B}(\sigma''_{B,\nu_{1}},\sigma''_{B,\nu_{2}})\le2\sqrt{\delta^{4}_{U,\nu_{1}}+\delta^{4}_{U,\nu_{2}}+\sqrt{2}(\delta^{2}_{U,\nu_{1}}+\delta^{2}_{U,\nu_{2}})}+2\sqrt{2}\delta^{2}_{U,\nu_{1+2}}.
\end{align}

Finally, using the triangle inequality, we have \eref{u-ineq}:
\balign{
L_{B}(\sigma'_{B,\nu_{1}},\sigma'_{B,\nu_{2}})\leq& L_{B}(\sigma'_{B,\nu_{1}},\sigma''_{B,\nu_{1}})+L_{B}(\sigma''_{B,\nu_{1}},\sigma''_{B,\nu_{2}})+L_{B}(\sigma''_{B,\nu_{2}},\sigma'_{B,\nu_{2}}) \nt \\
\leq& 2\sqrt{\delta^{4}_{U,\nu_{1}}+\delta^{4}_{U,\nu_{2}}+\sqrt{2}(\delta^{2}_{U,\nu_{1}}+\delta^{2}_{U,\nu_{2}})}+2\sqrt{2}\delta^{2}_{U,\nu_{1+2}}+\delta^{2}_{U,\nu_{1}}+\delta^{2}_{U,\nu_{2}}
}
}

\section{Complete proof of Eq.(\ref{tradeoff})}
In the outline of the proof of \eqref{tradeoff} in the body of the text, we postponed the proof of Eqs.~\eqref{lemma1}, \eqref{lemma2}, \eqref{s1}, and \eqref{s2}.
Because we have shown the proof of Eqs.~\eqref{lemma1}, \eqref{lemma2}, here, we show the proof of Eqs.~\eqref{s1} and \eqref{s2}, which completes the proof of \eqref{tradeoff}.

\bpf{[Proof of \eref{s1}]
First, we define the following energy differences:
\begin{align}
\Delta_{\max}&:=\Tr[H_{E}(\sigma'_{E,\max}-\sigma_{E})]=\Tr[H_{S}(\rho_{S,\max}-\rho'_{S,\max})],\\
\Delta_{\min}&:=\Tr[H_{E}(\sigma'_{E,\min}-\sigma_{E})]=\Tr[H_{S}(\rho_{S,\min}-\rho'_{S,\min})],\\
\Delta_{U,\max}&:=\Tr[H_{S}(\rho_{S,\max}-U_{S}\rho_{S,\max}U^{\dagger}_{S})],\\
\Delta_{U,\min}&:=\Tr[H_{S}(\rho_{S,\min}-U_{S}\rho_{S,\min}U^{\dagger}_{S})].
\end{align}
Then,
\begin{align}
\Delta&=|\Delta_{\max}-\Delta_{\min}|,\label{8.22.1}\\
\|[H_S,U_S]\|_{}&=\max\left(|\Delta_{U,\max}|, |\Delta_{U,\min}|\right),\label{8.22.2}\\
|\Delta_{i}-\Delta_{U,i}|&\le\|\rho'_{S,i}-U_{S}\rho_{S,i}U^{\dagger}_{S}\|_{1}\|H_{S}\|_{}\nonumber\\
&\leq 2\sqrt{2}L_{B}(\rho'_{S,i},U_{S}\rho_{S,i}U^{\dagger}_{S})\|H_{S}\|_{}\le2\sqrt{2}\delta^{2}_{U}\|H_{S}\|_{}.\label{8.22.3}
\end{align}
where $i\in \{ \max,\min\}$, $\| X\|_1:=\Tr \sqrt{X^\dagger X}$, and we used $\|\rho-\sigma\|_{1}\le2\sqrt{1-F^2(\rho,\sigma)}\le2\sqrt{2}L_{B}(\rho,\sigma)$ \cite{hayashi} in the fourth line.

Because $\Tr[H_{S}(U_{S}\hat{1}_{S}U^{\dagger}_{S}-\hat{1}_{S})]=0$, the signs of $\Tr[H_{S}(U_{S}\rho_{S,\max}U^{\dagger}_{S}-\rho_{S,\max})]$ and $\Tr[H_{S}(U_{S}\rho_{S,\min}U^{\dagger}_{S}-\rho_{S,\min})]$ are different. Therefore, we have
\begin{align}
\max\left(|\Delta_{U,\max}|, |\Delta_{U,\min}|\right)\le|\Delta_{U,\max}-\Delta_{U,\min}|\label{8.22.4}
\end{align}
Combining \eqref{8.22.1}--\eqref{8.22.4}, we obtain the desired inequality:
\begin{align}
\|[H_S,U_S]\|_{}&=\max\left(|\Delta_{U,\max}|, |\Delta_{U,\min}|\right)\nonumber\\
&\le|\Delta_{U,\max}-\Delta_{U,\min}|\nonumber\\
&\le|\Delta_{\max}-\Delta_{\min}|+4\sqrt{2}\delta^{2}_{U}\|H_{S}\|_{}\nonumber\\
&=\Delta+4\sqrt{2}\delta^{2}_{U}\|H_{S}\|_{}.\label{12.19.1}
\end{align}
}

\bpf{[Proof of \eqref{s2}]
The energy-preserving property of $\Lambda_{SE}$ gives
\begin{align}
\delta^{2}_{S}(\rho_{S,\nu})+\delta^{2}_{E}(\sigma_{E})=\delta^{2}_{S}(\rho'_{S,\nu})+\delta^{2}_{E}(\sigma'_{E,\nu})+2{\rm Cov}_{SE}(\Lambda_{SE}(\rho_{S,\nu}\otimes\sigma_{E})),
\end{align}
where $\delta_{S}(\rho)$ is the standard deviation of the energy in $\rho$, and ${\rm Cov}_{SE}(\sigma)$ is the energy covariance of the state of $\sigma$ on $SE$.
Because $-\delta_{S}(\rho'_{S,\nu})\delta_{E}(\sigma'_{E,\nu})\le {\rm Cov}_{SE}(\Lambda_{SE}(\rho_{S,\nu}\otimes\sigma_{E}))$ (this is a basic feature of the covariance), we obtain
\begin{align}
\delta_{E}(\sigma'_{E,\nu})-\delta_{S}(\rho'_{S,\nu})\leq \sqrt{\delta^{2}_{S}(\rho'_{S,\nu})+\delta^{2}_{E}(\sigma'_{E,\nu})-2\delta_{S}(\rho'_{S,\nu})\delta_{E}(\sigma'_{E,\nu})} \leq  \sqrt{\delta^2_{E}(\sigma_{E})+\delta^2_{S}(\rho_{S,\nu})} \leq \delta_{E}(\sigma_{E})+\delta_{S}(\rho_{S,\nu}).
\end{align}
Because the standard deviation of the energy in $S$ is always smaller than $\|H_{S}\|_{}/2$, we obtain \eref{s2}.
}

\newcommand{\sgme}{\sigma_E}

\section{Complete proof of Eq.(\ref{tradeoff2})}
Here, we demonstrate the complete proof of \eqref{tradeoff2}.
Without loss of generality, we can assume that the expression of the initial state of the external system $\sgme:= \sum_j p_j |\phi_{E,j} \rangle \langle \phi_{E,j} |$ given in \eref{sigmaec} satisfies $\delta_{EQ}=\sum_{j}p_{j}\delta_{E}(\braket{\phi_{E,j}})$.
Using the expression, we define
\begin{align}
\begin{split}
\sigma_{E, j}' &:= {\rm Tr}_S \left[ \Lambda_{SE} (\rho_{S} \otimes |\phi_{E,j}\rangle\langle \phi_{E,j}| )  \right] \, ,  \\
\sigma_{E,\nu}' &:= {\rm Tr}_S \left[ \Lambda_{SE} (\rho_{S,\nu} \otimes \sgme )  \right] \, , \\
\sigma_{E, ({\nu},j)}' &:= {\rm Tr}_S \left[ \Lambda_{SE} (\rho_{S,\nu} \otimes 
|\phi_{E,j}\rangle\langle \phi_{E,j}| )  \right]
\, , 
\end{split}
\end{align}
where $\nu$ takes $``{\rm max}"$ or $``{\rm min}"$, as in (\ref{maxmin}). 
$\rho_{S, j}', \rho_{S,\nu}'$, and $\rho_{S, ({\nu},j)}'$ are defined in a similar manner. 
We consider the degree of closeness to the unitary operator $U_S$, which is quantified as
\begin{align}
\begin{split}
\delta_{U} (\rho_S ) &= L_B \left( \rho_{S}' , U_S( \rho_S\otimes \sgme ) U_S^{\dagger}) \right)\, , \\
\delta_{U,j} (\rho_S ) &:= L_B \left( \rho_{S, j}' , U_S( \rho_S\otimes|\phi_{E,j}\rangle\langle \phi_{E,j}|) U_S^{\dagger}) \right)\, .
\end{split}
\end{align}

In this proof, we first follow the derivation of \eref{tradeoff} for each $|\phi_{E,j} \rangle \langle \phi_{E,j} |$, and we then sum it with $j$.
The inequality (\ref{middle}) for $\sigma_{E, ({\rm max},j)}'$ and $\sigma_{E, ({\rm min},j)}'$ reads
\begin{align}
\Delta_j \le M_j (2 \delta_{E,j}'' + \Delta_j ) \, ,\label{2-1}
\end{align}
where $M_{j}:=L_{B}(\sigma_{E, ({\rm max},j)}',\sigma_{E, ({\rm min},j)}')$, $\Delta_j :=\abs{ {\rm Tr} [ H_{E} (\sigma_{E,({\rm max},j)}'  - \sigma_{E,({\rm min},j)}' )]}$ and 
$\delta_{E,j}''$ is the larger of $\delta_E(\sigma_{E,({\rm max},j)}')$ or $\delta_E(\sigma_{E,({\rm min},j)}')$. 
Combining \eqref{s1}, \eqref{s2}, \eqref{2-1}, $\Delta_{j}\le \|H_{S}\|_{}$ and $\delta_U\le1/64$, we obtain
\begin{align}
\|[H_S,U_S]\|_{}&\le\Delta+\delta_U\|H_{S}\|_{} \nt \\
&=\sum_{j}p_{j}\Delta_{j}+\delta_U\|H_{S}\|_{}\nonumber\\
&\le\sum_{j}p_{j}M_j (2 \delta_{E,j}'' + \Delta_j )+\delta_U\|H_{S}\|_{}\nonumber\\
&\le\sum_{j}p_{j}M_j (2 \delta_{E,j}+3\|H_{S}\|_{})+\delta_U\|H_{S}\|_{}\nonumber\\
&\leq \sqrt{\sum_{j}p_{j}M^{2}_{j}}\left(2\sqrt{\sum_{j}p_{j}\delta^{2}_{E,j}}+3\|H_{S}\|_{}\right)+\delta_U\|H_{S}\|_{}\nonumber\\
&=\sqrt{\sum_{j}p_{j}M^{2}_{j}}\left(2\delta_{EQ}+3\|H_{S}\|_{}\right)+\delta_U\|H_{S}\|_{}\label{2-2}
\end{align}
where we used \eqref{s2} and $\Delta_{j}\le \|H_{S}\|_{}$ in the fourth line, and we used the Schwarz inequality in the fifth line.
If $\sgme$ is a pure state, then $\sqrt{\sum_{j}p_{j}M^{2}_{j}}=L_{B}(\sigma_{E, {\rm max})}',\sigma_{E, {\rm min})}')\leq 4\delta_U$, and the above inequality is transformed into $\calE_{\max}\leq 8\delta_U\delta_{EQ}+13\delta_U\|H_{S}\|_{}$, for which we have seen a similar relation in the derivation of \eqref{tradeoff}.

Below, we investigate the upper bound of $\sum_{j}p_{j}M^{2}_{j}$ in the form of $\sum_{j}p_{j}M^{2}_{j}\le a\delta^2_{U}$.
We first show the bound of $M_j$:
\begin{align}
M_j &\le 3\sqrt{2}\delta_{U,j} (\rho_{S,{\rm max}})+ 3\sqrt{2}\delta_{U,j} (\rho_{S,{\rm min}})  + 2\sqrt{2}\delta_{U,j}(\rho_{S,\max+\min}) \, ,  \label{eq2}
\end{align}
where we defined $\ket{\rho_{S,\max+\min}}:=(\ket{\rho_{S,\max}}+\ket{\rho_{S,\min}})/\sqrt{2}$.

\bpf{[Proof of \eqref{eq2}]
We first consider the case where $\Lambda_{SE}$ is a unitary dynamics $\Lambda_{SE}$.
In this case, setting $S$ and $E$ to $A$ and $B$, and substituting $\ket{\psi_{E,j}}\bra{\psi_{E,j}}$ into $\sigma_{E}$ in Theorem \ref{unify}, we obtain the desired inequality
\begin{align}
M_{j}&\le2\sqrt{\delta^{4}_{U,j}(\rho_{S,\max})+\delta^{4}_{U,j}(\rho_{S,\min})+\sqrt{2}(\delta^{2}_{U,j}(\rho_{S,\max})+\delta^{2}_{U,j}(\rho_{S,\min}))}+2\sqrt{2}\delta^{2}_{U,j}(\rho_{S,\max+\min})+\delta^{2}_{U,j}(\rho_{S,\max})+\delta^{2}_{U,j}(\rho_{S,\min})\nonumber\\
&\leq \left(2\sqrt{1+\sqrt{2}}+1\right)\delta_{U,j} (\rho_{S,{\rm max}})+\left(2\sqrt{1+\sqrt{2}}+1\right)\delta_{U,j} (\rho_{S,{\rm min}})+2\sqrt{2}\delta_{U,j}(\rho_{S,\max+\min})\nonumber\\
&\le3\sqrt{2}\delta_{U,j} (\rho_{S,{\rm max}})+ 3\sqrt{2}\delta_{U,j} (\rho_{S,{\rm min}})  + 2\sqrt{2}\delta_{U,j}(\rho_{S,\max+\min}).
\end{align}
In the second line, we used $0\le\delta_{U,j}(\rho_{S,\nu})\le1$ $(\nu=\max,\min,\max+\min)$. 
We can handle the general case where $\Lambda_{SE}$ is a general CPTP map in a similar manner to that of \eqref{lemma1}.
}

Next, we introduce the following inequality for any pure state $\rho_{S}$
\begin{eqnarray}
\sum_{j} p_j \delta_{U,j}^4 (\rho_{S} ) \le 2 \delta_U^4 (\rho_S)\, . \label{eq1}
\end{eqnarray}
which follows from
\balign{
1-2\delta^{4}_{U}(\rho_{S})&\le(1-\delta^{4}_{U}(\rho_{S}))^2=\bra{\rho''_{S}}\rho'_{S}\ket{\rho''_{S}}
=\sum_{j}p_{j}\bra{\rho''_{S}}\rho'_{S,j}\ket{\rho''_{S}}=\sum_{j}p_{j}(1-\delta^{4}_{U}(\rho_{S,j}))^2
\le\sum_{j}p_{j}(1-\delta^{4}_{U}(\rho_{S,j})).
}
Here, $\ket{\rho''_{S}}$ is the vector representation of $U_{S}\rho_{S}U^{\dagger}_{S}$.
Combining \eqref{eq1}, \eqref{eq2}, and $(A+B+C)^2 \le 3(A^2 + B^2 + C^2)$, we arrive at the desired upper bound
\begin{align}
\sum_{j}p_{j}M^{2}_{j}&\le3\sum_{j}p_{j}\left(18\delta^{2}_{U,j} (\rho_{S,{\rm max}})+ 18\delta^{2}_{U,j} (\rho_{S,{\rm min}})  + 8 \delta^{2}_{U,j}(\rho_{S,\max+\min})\right)\nonumber\\
&\le54\sqrt{\sum_{j}p_{j}\delta^{4}_{U,j} (\rho_{S,{\rm max}})}+54\sqrt{\sum_{j}p_{j}\delta^{4}_{U,j} (\rho_{S,{\rm min}})}+24\sqrt{\sum_{j}p_{j}\delta^{4}_{U,j} (\rho_{S,\max+\min})}\nonumber\\
&\le132\sqrt{2}\delta^{2}_U. \label{2-3}
\end{align}

Substituting \eqref{2-3} into \eqref{2-2} and noting $\delta_U\le\|[H_S,U_S]\|_{}/(64\|H_{S}\|_{})$, we obtain our main result \eqref{tradeoff2}:
\begin{align}
\|[H_S,U_S]\|_{}\le\frac{2\sqrt{132\sqrt{2}}}{63-3\sqrt{132\sqrt{2}}}64\delta_U\delta_{EQ}\leq 81\delta_U\delta_{EQ}.\label{8.22.5}
\end{align}

\section{Jayes-Cummings model}
We consider the Jayes-Cummings Hamiltonian
\begin{eqnarray}
H &=& \epsilon \sigma_z + \lambda (\sigma_+ b  + b^{\dagger} \sigma_- ) + 2 \epsilon  b^{\dagger} b \, .
\end{eqnarray}
We derive the following time-dependent Hamiltonian $\tilde{H}_{S} (t)$ for the initial state $|\psi_{\rm ini}\rangle  = |\psi_S\rangle \otimes |\alpha \rangle$ with the limit of $\lambda\to +0$, keeping $\lambda \alpha$ constant:
\begin{eqnarray}
\tilde{H}_{S} (t) &=& \epsilon \sigma_z + \lambda \alpha ( \sigma_+ e^{-i 2 \epsilon t} + \sigma_- e^{i 2\epsilon t} ) \, .
\label{tildehs}
\end{eqnarray}

We start with the expression of the bosonic operator at time $t$ in the form
\begin{eqnarray}
b(t) &=& e^{-2i \epsilon t } b - i \lambda \int_0^t ds \, e^{-2i \epsilon (t-s)} \sigma_- (s) \, . \label{b-exp}
\end{eqnarray}
The equations of motion for the spin operators are given by
\begin{eqnarray}
{\partial \sigma_- \over \partial t} = -2i \epsilon \sigma_- + i \lambda \sigma_z b  \, , ~~&&~~ {\partial \sigma_z \over \partial t} = -2i \lambda \left( \sigma_+ b - b^{\dagger} \sigma_- \right) \, . 
\end{eqnarray}
By substituting Eq.(\ref{b-exp}) into these equations, we have 
\begin{align}
{\partial \sigma_- \over \partial t} &= -2i \epsilon \sigma_- (t) 
+ i \lambda \left\{ e^{-2i\epsilon t} \sigma_z (t) b  - i \lambda \int_0^t ds \, e^{-2i\epsilon (t-s) } 
\sigma_z (t) \sigma_- (s)  \right\} \, ,  \\
{\partial \sigma_z \over \partial t} &= -2i \lambda 
\left( e^{-2i \epsilon t} \sigma_+ (t) b  - e^{2i\epsilon t }b^{\dagger}  \sigma_-(t) \right) 
- 2 \lambda^2 \int_0^t ds \, e^{2i\epsilon (t-s) } 
\left\{ 
\sigma_+ (t) \sigma_- (s) + \sigma_+ (s) \sigma_- (t) 
\right\} \, . 
\end{align}
Now, we consider the average over the initial state $|\psi_{\rm ini}\rangle  = |\psi_S\rangle \otimes |\alpha \rangle$. Noting the relation $b | \alpha \rangle = \alpha | \alpha \rangle$, we find the following expression:
\begin{eqnarray}
{\partial \langle \sigma_- \rangle \over \partial t} &=& -2i \epsilon \langle \sigma_- (t) \rangle + i \lambda \alpha e^{-2i\epsilon t} \langle \sigma_z (t) \rangle +  \lambda^2 \int_0^t ds \, e^{-2i\epsilon (t-s) } \langle \sigma_z (t) \sigma_- (s)  \rangle \, , \label{av1} \\
{\partial \langle \sigma_z \rangle \over \partial t} &=& -2i \left( \lambda \alpha e^{-2i \epsilon t} \langle \sigma_+ (t) \rangle   
- e^{2i\epsilon t } \lambda \alpha^{\ast}  \langle \sigma_-(t) \rangle \right) - 2 \lambda^2 \int_0^t ds \, e^{2i\epsilon (t-s) } 
\left\{ 
\langle \sigma_+ (t) \sigma_- (s) \rangle + \langle \sigma_+ (s) \sigma_- (t) \rangle  
\right\} \, ,  \label{av2}
\end{eqnarray}
where $\langle ...\rangle = \langle \psi_{\rm ini} | ... | \psi_{\rm ini} \rangle$. Note here the following expression
\begin{eqnarray}
\langle \sigma_{a} (t) \sigma_b (s) \rangle &=& \langle \psi_{\rm ini}| U_t^{\dagger} \sigma_a U_t U_s^{\dagger} \sigma_b U_s  | \psi_{\rm ini} \rangle  \, , \label{ss1}
\end{eqnarray}
where $U_t$ is the time-evolution operator. Note that the time-evolution operator can be expanded as
\begin{eqnarray}
U_t &=& e^{-i H_0 t} + e^{-i H_0 t} \int_0^t \lambda (\tilde{\sigma}_+ (u) \tilde{b} (u) +
 \tilde{b}^{\dagger} (u) \tilde{\sigma}_- (u)  ) +\cdots \, . \label{ss2}
 \end{eqnarray}
We used the interaction picture, i.e., $\tilde{A} (u) := e^{i H_0 u} A e^{-i H_0 u}$, where $H_0 := \epsilon \sigma_z + 2\epsilon b^{\dagger} b $. From the expressions (\ref{ss1})
and (\ref{ss2}), we recognize that the quantity $\langle \sigma_{a} (t) \sigma_b (s) \rangle $ is a function of $\alpha \lambda$. 

Now, we impose the following condition
\begin{eqnarray}
\lambda\to +0  ~~ {\rm with}~~\alpha \lambda ={\rm constant} \, . \label{cond}
\end{eqnarray}
From Eqs.(\ref{av1}) and (\ref{av2}) as well as the observation that the quantity $\langle \sigma_{a} (t) \sigma_b (s) \rangle $ is a function of $\alpha \lambda$, this condition justifies the following approximation for the equations of motion for spin variables:
\begin{eqnarray}
{\partial \langle \sigma_- \rangle \over \partial t} &\sim& -2i \epsilon \langle \sigma_- (t) \rangle + i \lambda \alpha e^{-2i\epsilon t} \langle \sigma_z (t) \rangle \, , \label{avv1} \\
{\partial \sigma_z \over \partial t} &\sim& -2i \left( \lambda \alpha e^{-2i \epsilon t} \langle \sigma_+ (t) \rangle   
- e^{2i\epsilon t } \lambda \alpha^{\ast}  \langle \sigma_-(t) \rangle \right)  \, ,  \label{avv2}
\end{eqnarray}
which is consistent with the description that has the desired time-dependent Hamiltonian $\tilde{H}_S(t)$.
We assumed that the time-integration terms in Eqs.(\ref{av1}) and (\ref{av2}) never diverge. Note that the equations (\ref{avv1}) and (\ref{avv2}) are exactly derived from the effective Hamiltonian
(\ref{tildehs}). 

\section{Results for the cases where $[H_S+H_E,e^{-iH\tau}]=0$ does not hold}

In this letter, we assume $[H_S+H_E,e^{-iH\tau}]=0$ as the energy conservation law.
This is the energy conservation excluding $H_{SE}$, and thus it does not hold in general.
Therefore, in this section, we give generalized versions of our results which are valid for the case where $H_S+H_E$ energy conservation is not satisfied.

Let us remove the assumption $[H_S+H_E,e^{-iH\tau}]=0$ from our setup.
(We do not make other changes. Therefore, we treat the time-independent $H_{SE}$. We can easily extend the results in this section to the case where $H_{SE}$ is time-dependent by substituting $U_{\mathrm{tot}}={\cal T}[\exp(-i\int^{\tau}_{0}H_S+H_E+H_{SE}(s)ds)]$ for $e^{-iH\tau}$.)
Then, we give the following inequalities with using $\chi:=\|[H_S+H_E,e^{-iH\tau}]\|$, which is an index of breaking of $H_S+H_E$ energy conservation:
\begin{theorem}\label{nonpreserving}
\begin{align}
\delta_{E}\delta_{U}&\ge\frac{\|[U_S,H_S]\|-\chi}{40},\label{smain1}\\
\delta_{EQ}\delta_{U}&\ge\frac{\|[U_S,H_S]\|-\chi}{81}\label{smain2}
\end{align}
for $\delta_{U}<\frac{\|[U_S,H_S]\|-\chi}{128\max\{\|H_S\|,\chi\}}$.
\end{theorem}
The quantity $\chi=\|[H_S+H_E,e^{-iH\tau}]\|$ is the maximum change in the expectation value of $H_S+H_E$:
\begin{align}
\|[H_S+H_E,e^{-iH\tau}]\|=\max_{\rho}|\mathrm{Tr}[(\rho-e^{-iH\tau}\rho e^{iH\tau})(H_S+H_E)]|,
\end{align}
where $\max_{\rho}$ is the maximization through $\rho$ on the composite system $SE$.
Thus, we can interpret $\chi$ as an indicator describing how the $H_S+H_E$-energy conservation breaks in the meaning of the expectation value.
As we have pointed out in the main text, $\|[U_S,H_S]\|$ is the maximum change of the expectation value of $H_S$ caused by the desired unitary dynamics $U_S$.
Therefore, Theorem \ref{nonpreserving} means that our uncertainty relations are qualitatively valid as long as the maximum change in the expectation value of $H_S+H_E$ is smaller than that of $H_S$.

We show the above theorem with using the following lemma:
\begin{lemma}
The following inequalities hold:
\begin{align}
\|[U_S,H_S]\| &\le \Delta+4\sqrt{2}\delta^{2}_{U}\|H_S\|+\chi,\label{s1'}\\
\tilde{\delta}_{E}&\le \delta_{E}+2\max\{\|H_S\|,\chi\}\label{s2'}
\end{align}
\end{lemma}
The inequality \eqref{s1'} is the inequality \eqref{s1} with $\chi$ in righthand side.
The inequality \eqref{s2'} is the inequality \eqref{s2} whose $\|H_S\|$ is substituted by $2\max\{\|H_S\|,\chi\}$.
In the proof of Theorem \ref{nonpreserving}, we will use \eqref{s1'} and \eqref{s2'} instead of \eqref{s1} and \eqref{s2}.
\begin{proof}
We firstly show \eqref{s1'}.
We define
\begin{align}
\Delta^{S}_{\max}&:=\Tr[H_{S}(\rho_{S,\max}-\rho'_{S,\max})],\\
\Delta^{S}_{\min}&:=\Tr[H_{S}(\rho_{S,\min}-\rho'_{S,\min})],
\end{align}
Then, clearly $|\Delta^{S}_{\max}-\Delta^{S}_{\min}|\le \Delta+\chi$.
In the same manner as the derivation of \eqref{12.19.1}, we obtain
\begin{align}
\|[H_S,U_S]\|\le|\Delta^{S}_{\max}-\Delta^{S}_{\min}|+4\sqrt{2}\delta^{2}_{U}\|H_S\|.
\end{align}
Therefore, we obtain \eqref{s1'}.

Next, we derive \eqref{s2'}.
We use the following important fact:$\\$
\textit{
Let us take an arbitrary positive operator $A$ and arbitrary unitary $U$.
When $\|[U,A]\|\le\chi$ holds,  the following inequality holds for an arbitrary state $\rho$:}
\begin{align}
|\delta^{2}_{A}(\rho)-\delta^{2}_{U^{\dagger}AU}(\rho)|\le\chi(2\delta_{A}(\rho)+\chi),\label{fact}
\end{align}
\textit{where $\delta_{A}(\rho)$ is the standard deviation of $A$ in $\rho$.}$\\$
(\textit{Proof of \eqref{fact}:}
Because of $\|[A,U]\|=\|A-U^{\dagger}AU\|$, the Hermitian $X:=A-U^{\dagger}AU$ satisfies $\|X\|\le\chi$.
With using $X$, we can express $\delta^{2}_{U^{\dagger}AU}$ as follows:
\begin{align}
\delta^{2}_{U^{\dagger}AU}(\rho)&=\left<(A-X)^2\right>_{\rho}-\left<A-X\right>^{2}_{\rho}\nonumber\\
&=\delta^{2}_{A}(\rho)-2{\rm Cov}_{A;X}(\rho)+\delta^{2}_{X}(\rho),
\end{align}
where ${\rm Cov}_{A;X}(\rho):=\frac{1}{2}\Tr[\rho(AX+XA)]-\left<A\right>_{\rho}\left<X\right>_{\rho}$.
Because of $\delta_{X}(\rho)\le\|X\|\le\chi$ and the quantum correlation coefficient is lower than or equal to $1$, we obtain
\begin{align}
|\delta^{2}_{U^{\dagger}AU}(\rho)-\delta^{2}_{A}(\rho)|\le2|{\rm Cov}_{A;X}(\rho)|+\delta^{2}_{X}(\rho)\le2\delta_{X}(\rho)\delta_{A}(\rho)+\delta^{2}_{X}(\rho)\le \chi(2\delta_{A}(\rho)+\chi).
\end{align}
\textit{(Proof end)})

Let us show \eqref{s2'}.
With using \eqref{fact}, we firstly show that the variances of $H_S+H_E$ in the initial and the final states are very close to each other.
The variance of $H_S+H_E$ in the initial state is $\delta^{2}_{S}(\rho_{S,\nu})+\delta^{2}_{E}(\sigma_{E})$, and corresponds to $\delta^{2}_{A}(\rho)$ in \eqref{fact}.
The variance of $H_S+H_E$ in the final state is $\delta^{2}_{S}(\rho'_{S,\nu})+\delta^{2}_{E}(\sigma'_{E,\nu})+2{\rm Cov}_{SE}(e^{-iH\tau}(\rho_{S,\nu}\otimes\sigma_{E})e^{iH\tau})$, and corresponds to $\delta^{2}_{U^{\dagger}AU}(\rho)$ in \eqref{fact}.
Substituting $H_S+H_E$, $e^{-iH\tau}$ and $\rho_{S,\nu}\otimes\sigma_{E}$ for $A$, $U$ and $\rho$ of \eqref{fact}, we obtain 
\begin{align}
\delta^{2}_{S}(\rho_{S,\nu})+\delta^{2}_{E}(\sigma_{E})\ge\delta^{2}_{S}(\rho'_{S,\nu})+\delta^{2}_{E}(\sigma'_{E,\nu})+2{\rm Cov}_{SE}(e^{-iH\tau}(\rho_{S,\nu}\otimes\sigma_{E})e^{iH\tau})-\chi(2\sqrt{\delta^{2}_{S}(\rho_{S,\nu})+\delta^{2}_{E}(\sigma_{E})}+\chi),
\end{align}
where $\delta_{S}(\rho)$ is the standard deviation of the energy in $\rho$, and ${\rm Cov}_{SE}(\sigma)$ is the energy covariance of the state of $\sigma$ on $SE$.
Because $-\delta_{S}(\rho'_{S,\nu})\delta_{E}(\sigma'_{E,\nu})\le {\rm Cov}_{SE}(e^{-iH\tau}(\rho_{S,\nu}\otimes\sigma_{E})e^{iH\tau})$ (this is a basic feature of the covariance) and $\delta_{S}(\rho)\le\|H_S\|/2$ for any $\rho$, we obtain
\begin{align}
\delta_{E}(\sigma'_{E,\nu})-\delta_{S}(\rho'_{S,\nu})&\leq \sqrt{\delta^{2}_{S}(\rho'_{S,\nu})+\delta^{2}_{E}(\sigma'_{E,\nu})-2\delta_{S}(\rho'_{S,\nu})\delta_{E}(\sigma'_{E,\nu})} \nonumber\\
&\leq  \sqrt{\delta^2_{E}(\sigma_{E})+\delta^2_{S}(\rho_{S,\nu})+\chi(2\sqrt{\delta^{2}_{S}(\rho_{S,\nu})+\delta^{2}_{E}(\sigma_{E})}+\chi)}\nonumber\\
&\le \delta_{E}+1.5\max\{\|H_S\|,\chi\}.
\end{align}
Therefore, we obtain \eqref{s2'}.
\end{proof}

Next, we prove Theorem \ref{nonpreserving}.

\begin{proof}
We firstly show the inequality \eqref{smain1}.
Note that the inequality \eqref{ul1-last} is shown by combining \eqref{lemma1}, \eqref{lemma2} and \eqref{s1}.
Because we do not use the assumption $[H_S+H_E,e^{-iH\tau}]=0$ in the proofs of the inequalities \eqref{lemma1} and \eqref{lemma2}, they are valid even when $[H_S+H_E,e^{-iH\tau}]\ne0$.
Thus, substituting \eqref{s1'} for \eqref{s1} in the derivation of \eqref{ul1-last}, we obtain
\begin{align}
\delta_{U}\tilde{\delta}_{E}\ge\frac{\|[H_S,U_S]\|-\chi}{20}.\label{12.19.2}
\end{align}
Combining \eqref{s2'}, \eqref{12.19.2} and $\delta_{U}<\frac{\|[U_S,H_S]\|-\chi}{128\max\{\|H_S\|,\chi\}}$, we obtain \eqref{smain1}.

Secondly, we show the inequality \eqref{smain2}.
Note that the inequality \eqref{tradeoff2} is shown by combining \eqref{2-2} and \eqref{2-3}.
The inequality \eqref{2-3} is derived from Theorem 1.
Because we do not use the assumption $[H_S+H_E,e^{-iH\tau}]=0$ in the proof of Theorem 1, the inequality \eqref{2-3} is valid even when $[H_S+H_E,e^{-iH\tau}]\ne0$.
In the same manner as the derivation of \eqref{2-2}, we derive the following inequality by combining \eqref{s1'}, \eqref{s2'}, \eqref{2-1}, $\Delta_{j}\le \|H_{S}\|_{}+\chi$ and $\delta_U\le1/128$:
\begin{align}
\|[H_S,U_S]\|-\chi\le\sqrt{\sum_{j}p_jM^{2}_{j}}(2\delta_{EQ}+4\max\{\|H_S\|,\chi\}+\chi)+\delta_{U}\|H_S\|\label{2-2'}
\end{align}
Combining \eqref{2-3} and \eqref{2-2'} and  $\delta_{U}<\frac{\|[U_S,H_S]\|-\chi}{128\max\{\|H_S\|,\chi\}}$, we obtain \eqref{smain2}.
\end{proof}

\section{Results with using entanglement fidelity}

As we have pointed out in the discussion in the main text, we can improve our inequality \eqref{tradeoff} with using entanglement fidelity.
In this section, we show the explicit expression of the improved inequality.

As the setup, we use the same one as the main text.
(We can use the energy-nonpreserving setup used in the previous section.
But for simplicity, we don't use it here.)
As the index of the degree of accuracy of approximating the desired unitary $U_S$, we introduce the entanglement-Bures length:
\begin{align}
\delta_{Ue}:=\max_{\rho_S}L_{e}(\Lambda_{S},U_{S},\rho_{S}).\label{errorinQ}
\end{align}
Here, $\Lambda_S( ...):=\Tr_{E}[\Lambda_{SE}(...\otimes\sigma_E)]$ is the CPTP-map that describes the true dynamics of $S$, and $L_{e}(\Lambda_{S},U_{S},\rho)$ is the entanglement Brues length, which is defined by the entanglement fidelity $F_{e}(\Lambda_{S},U_{S},\rho_{S})$ \cite{hayashi} as follows:
\begin{align}
L_{e}(\Lambda_{S},U_{S},\rho_{S})&:=\arccos F_{e}(\Lambda_{S},U_S,\rho_{S}),\label{uerror}\\
F_{e}(\Lambda_{S},U_S,\rho_{S})&:=\sqrt{\bra{\psi}_{SR} (1_{R}\otimes \Lambda_{U^{\dagger}_{S}}\circ\Lambda_{S}(\ket{\psi}_{SR} \bra{\psi}_{SR}))\ket{\psi}_{SR}},
\end{align}
where $R$ is a reference system, and $\ket{\psi}_{SR}$ is the purification of $\rho_{S}$ on $SR$, and $\Lambda_{U^{\dagger}_{S}}(\rho):=U^{\dagger}_{S}\rho U_{S}$. The symbol $a\circ b$ means that we successively perform the operation $a$ after $b$. Then, we obtain the relation:
\begin{align}
\delta_{Ue}\delta_{E}\ge\frac{\|[H_S,U_S]\|}{8}\label{8unc}
\end{align}
for $\delta_{Ue}\le\frac{\|[H_S,U_S]\|}{16\|H_S\|}$. The details will be reported elsewhere.

\end{widetext}


\begin{thebibliography}{00}
\bibitem{footnoteA}
We can easily extend our results to the case that $\Lambda_{SE}$ is a general energy preserving map.
In that case, we impose the following condition as the $H_S+H_E$-preserving law:
\begin{align}
\Tr[(H_S+H_E)(\rho_{SE}-\Lambda_{SE}(\rho_{SE}))]=0,
\end{align}
for any state $\rho_{SE}$ on the composite system $SE$.
Note that this extended version of our results includes the case with an arbitrary time-dependent interaction Hamiltonian, as long as this commutation relation is satisfied. Here, time-dependence in $H_{SE}$ stands for the switch on and off of the connection between (S) and (E).
Also, we can extend our results to the transient case that the final Hamiltonian of (S) is different from the initial one with using the controller trick in \cite{HO}.
\bibitem{footnoteC}
This result also can be extended to the case of general CPTP-map whose energy-preserving is weakly broken.
\end{thebibliography}
\end{document}